\documentclass[journal]{IEEEtran}

\usepackage{algpseudocode,algorithm}
\usepackage{amsmath,amsfonts}
\usepackage[pdftex]{graphicx}
\usepackage{bm}
\usepackage{multicol, blindtext}

\usepackage{etoolbox}
\patchcmd{\maketitle}{}{}{}{}

\begin{document}
%
\title{The Design and Deployment of an End-to-end IoT Infrastructure for the Natural Environment}
%
%
\author{Vatsala Nundloll, Barry Porter, Gordon Blair\thanks{Manuscript submitted October 15, 2018}\thanks{V. Nundloll, B. Porter, G. Blair and G. Dean are at Lancaster University, UK (email: vatsala@lancaster.ac.uk, barry.porter@comp.lancs.ac.uk, g.s.blair@googlemail.com, g.dean1@lancaster.ac.uk)}, Jack Cosby, Bridget Emmett\thanks{J. Cosby and B.Emmett are at Centre for Ecology and Hydrology Bangor, UK (email: jaccos@ceh.ac.uk, bae@ceh.ac.uk)}, Ben Winterbourn\thanks{B. Winterbourn was at Center of Ecology and Hydrology Bangor, UK (email: benwinterbourn@gmail.com)}, Graham Dean, Philip Beattie\thanks{P. Beattie was at Lancaster University, UK (email: p.beattie@lancaster.ac.uk)}, Rory Shaw\thanks{R. Shaw was at University of Bangor, UK (email: rory.shaw@bangor.ac.uk)}, Davey Jones, Dave Chadwick\thanks{D. Jones and D. Chadwick are at University of Bangor, UK (email: d.jones@bangor.ac.uk, d.chadwick@bangor.ac.uk)}, Mike Brown\thanks{M. Brown is at Centre for Ecology and Hydrology Lancaster, UK (email: mjbr@ceh.ac.uk)}, Wayne Shelley\thanks{W. Shelley is at British Geological Survey, UK (email:wael@bgs.ac.uk)}, Izhar Ullah\thanks{I. Ullah was at Lancaster University, UK (email:izhar81pk@gmail.com)}}

\markboth{IEEE Internet of Things Journal,~Vol.~V, No.~N, Publication Date: YYYY}%
{Nundloll \MakeLowercase{\textit{et al.}}: The Design and Deployment of an End-to-end IoT Infrastructure for the Natural Environment}
\maketitle

\begin{abstract}
Internet of Things (IoT) systems have seen recent growth in popularity for city and home environments. We report on the design, deployment and use of IoT infrastructure for environmental monitoring and management. Working closely with hydrologists, soil scientists and animal behaviour scientists, we successfully deployed and utilised a system to deliver integrated information across these two fields in the first such example of real-time multi-dimensional environmental science. We describe the design of this system, its requirements and operational effectiveness for hydrological, soil and ethological scientists, and our experiences from building, maintaining and using the deployment at a remote site in difficult conditions. Based on this experience we discuss key future work for the IoT community when working in these kinds of environmental deployments.
\end{abstract}

\begin{IEEEkeywords}
Wireless sensor networks, Internet of Things, environmental science, experiences
\end{IEEEkeywords}

\section{Introduction}
\label{sec:intro}


The Internet of Things (IoT) represents the next major step for the Internet as it evolves from a communication substrate that connects computers to one that connects and embraces everyday objects (things). This has the potential to revolutionize many different sectors of the economy and society more generally, e.g. enabling smart cities, smart transport systems, intelligent management of energy supplies, etc, all enabled by data collection from sensors. Most research in the Internet of Things has been carried out in cities and urban areas more generally (see Section \ref{sec:related}).

We see great potential, and indeed in many ways an even bigger potential for transformation in applying Internet of Things concepts to rural areas to both gain a better understanding of the natural environment and to manage this environment through appropriate interventions. However, the deployment of Internet of Things technology in the natural environment poses new challenges around, for example, limitations of power and Internet connectivity, and hostile operational conditions.

The focus of this paper is on the Environmental IoT project, a first attempt to instrument and manage an environmental catchment in all its facets, across different geographical locations and at all its scales. We also take this a stage further, building on our experiences in the Environmental Virtual Observatory pilot project \cite{elkhatib2013EUH}. This project utilised cloud computing to offer a shared repository of data, models and other tools and artifacts to allow a range of stakeholders to visualise data, to run models, to feed data into models, to deal with uncertainty in models, and to discuss the results with other stakeholders and communities. The Internet of Things and cloud computing are strongly complementary technologies that, when combined, can provide a complete, end-to-end technological infrastructure for a step change in our understanding of complex environmental factors.

This research has the potential to have major impact on many aspects of rural life – on farmers and associated agricultural businesses, the water industry, tourists and tourism related businesses, and society more generally. This has the potential to completely transform these associated businesses, enabling critical areas such as integrated land and water management, coastal zone protection and precision agriculture. We are particularly interested though in new kinds of science, and associated management strategies that stem from bringing real-time data sets together and observing related inter-dependencies.

Current practices in the environmental and earth sciences areas focus heavily on standalone data logger systems, with some early initiatives starting to embrace wireless sensor networks for environmental modelling (e.g., \cite{delin2005environmental,martinez2004ESN,martinez2006DSN,hart2015iot}). These initiatives though tend to focus on particular environmental facets at particular scales, e.g. focusing on habitat monitoring \cite{moreno2011TWN,szewczyk2004HMS}, glaciology \cite{martinez2009DWS}, permafrost \cite{Hasler2008WirelessSN} and volcanoes \cite{allen2006volcano}. While wireless networks have become quite sophisticated in many cases, they need a step change in their interoperability, scope and usability to become functioning Environmental IoT's, which also need to encompass widespread deployment of spatially distributed devices with embedded identification, sensing and/or actuation capabilities \cite{Miorandi20121497}. The development of a full Environmental IoT would also provide analytical tools to understand the functioning of natural systems based on real-time networks of sensors deployed widely across the landscape. Combined with other existing environmental data (maps of geology, topography, soils, etc.) and model outputs (rain and flood forecasts, etc) the Environmental IoT can provide the basis for decision and support systems for adaptive management of natural resources and for raising alerts. One of the major factors in the slow uptake of the technology from the lab to the field is because most of these systems rely on nonstandard, custom-designed elements that need specialised expertise \cite{hart2015iot}.

This paper reports in detail on experiences from the design and live deployment of an Environmental IoT targeting specifically local and regional environmental applications (hillslope to river catchment) using inexpensive off the shelf technologies, and deploying the system for a particular environmental issue: flood and pollution monitoring and alerts in rural environments. Our central hypothesis is that our combination of IoT technology coupled with Cloud Computing enables a paradigm shift in our understanding and management of the natural environment, especially related to understanding ecosystem inter-dependencies, in times of unprecedented environmental change.

The key contribution of this paper is an evaluation of this hypothesis in terms of technological, scientific and methodological aspects:

\begin{itemize}
    \item Technological: the presentation of a complete end-to-end systems architecture embracing cloud and IoT infrastructure and designed to operate effectively in the target operational environment;
    \item Scientific: the assessment of the potential impact of IoT / cloud technology on environmental and earth sciences;
    \item Methodological: a reflection on the methodological approaches required to facilitate the desired cross-disciplinary conversation to develop and utilise an Environmental IoT infrastructure.
\end{itemize}

The paper is structured as follows. Section \ref{sec:deployment} discusses the target deployment environment for the Environmental IoT infrastructure, highlighting the key challenges being addressed by the infrastructure. Section \ref{sec:related} then provides a more in-depth assessment of related work, supporting our view that the application of IoT principles in the natural environment has been limited in scope and vision. Section \ref{sec:approach} presents in detail the design and deployment of the Environmental IoT platform, and section \ref{sec:evaluation} then discusses experiences of this work. Section \ref{sec:discussion} provides a more high level discussion around the technological, scientific and methodological aspects discussed above. Finally, section \ref{sec:conclusion} presents concluding remarks.

\section{The Target Deployment Environment}
\label{sec:deployment}

\begin{figure}[!t]
\includegraphics[width=2.7in]{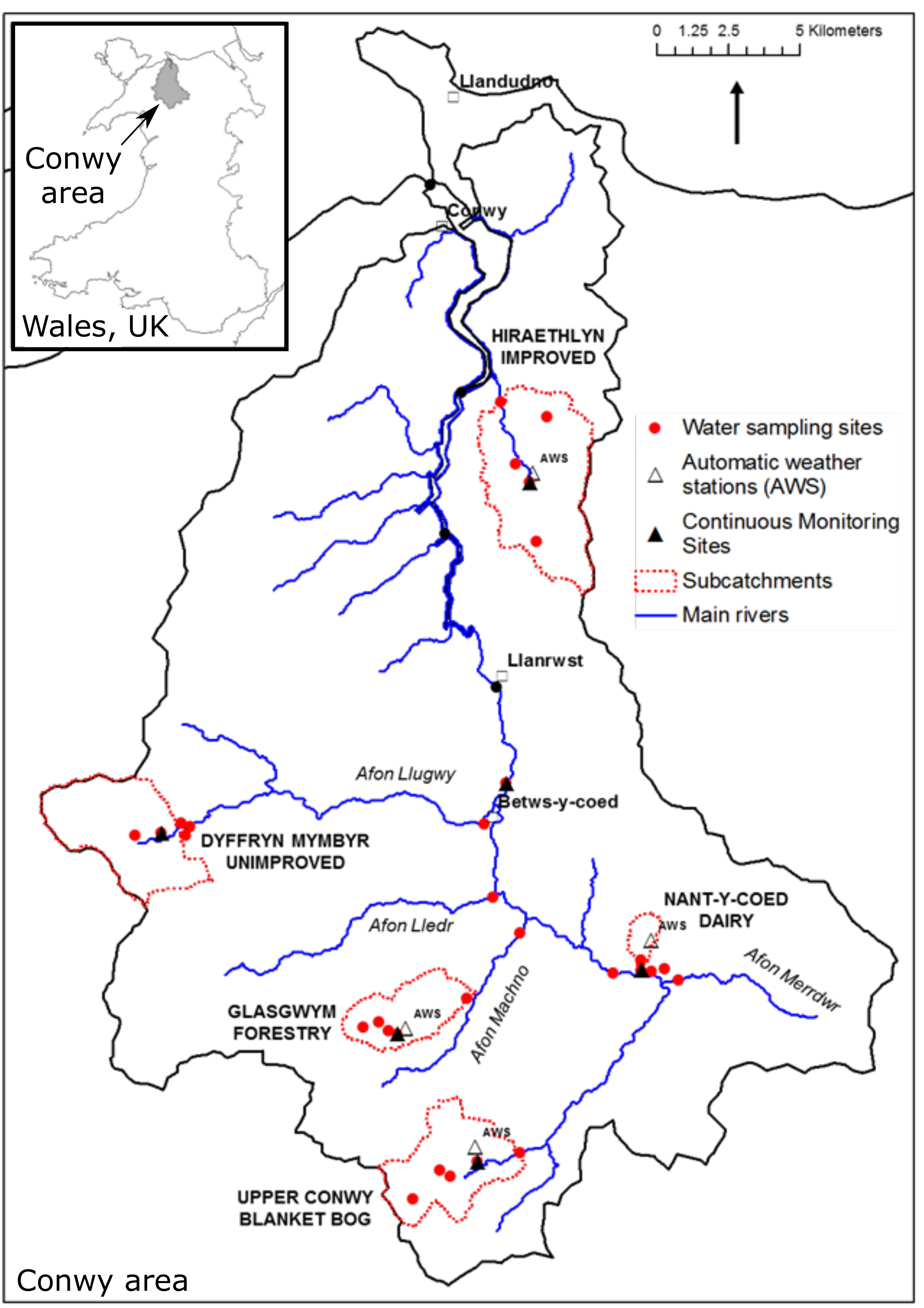}
\caption{Wide-area view of the geographical region of our deployment. Areas marked in dotted lines are those actively studied by the environmental science team using various types of traditional (non-IoT) measurement equipment. The primary site used for our IoT project is that marked `Hiraethlyn Improved'.}
\label{fig:region}
\end{figure}

We focus on one specific geographic region around Conwy in North Wales (see Fig.~\ref{fig:region}), typical of many rural areas supporting important industries including agriculture, forestry, tourism and fishing but facing huge challenges brought about by climate change and conflicting demands on land/ water resources. The Conwy river rises in the Snowdonia National Park, in the Migneint range, which has high rainfall providing water resources and important carbon storage and biodiversity in peatland habitats. The Conwy flows through intensive agricultural land out to an estuary important for the shellfish industry, conservation and tourism. The lower Conwy was recently in the national news due to intensive flooding of several larger towns. Such extreme events are consistent with climate change predictions. The major shellfisheries and tourist beaches close to the Conwy estuary are continually subject to contamination and risk of closure or loss of blue flag status due to microbial pollution from rural wastewater treatment plants and agricultural runoff. These pollution events are becoming increasingly frequent and are also exacerbated by floods, storms in general and periods of drought.

One of the main challenges in the Conwy and many other catchments / landscapes is the potential conflict arising from the needs of different industries, e.g. agriculture, water, tourism and urban development and the need for decision support tools to future-proof against climate change. GDP associated with Wales and the UK indicate all are important to local and national economies to varying degrees, and balancing their requirements is a major challenge for both government and industry alike. In the past 'silo management' has often resulted in a development by one industry negatively impacting on a second (e.g. intensification of agriculture reducing water quality) and there is a move towards more holistic consideration of these challenges (cf. Natural Environment White Paper \cite{newp}). However, a major problem identified in delivering this vision is the fragmented and inaccessible nature of the resources (i.e. data, models, tools) needed for a more integrated approach.

This problem is illustrated by the following scenario:

\begin{itemize}
\item[A]\textit{Representatives from the Conwy shellfish industry contacted Prof Emmett to ask why their shellfish products had just been rejected by Europe based on their high microbiological pollutant load and risk to public health. The industry wanted to know where the pollution was coming from, what caused the pollution to occur in the last month, and what safeguards they could put in place for the future. The professor used the Environmental IoT to identify any recent anomalous events in the Conwy catchment, which she suspected was the source of the problem. She determined that the recent movement of livestock into lowland areas, combined with high rainfall and spring tides, had caused a significant transfer of nutrients and faecal bacteria into coastal waters. She advised that the use of a mobile app would allow them to predict when these events might happen in the future and when their waters would be safe to harvest shellfish again.}
\end{itemize}

An Environmental IoT provides a major opportunity to fill this gap bringing together data and assets from across different domains (soil, water, plant ecology, animals) and organisations (e.g. public, regulatory, industry) to enable integrated problem solving for the agriculture and water industries but with clear potential to link to other critical industries in the future.

\section{Related Work}
\label{sec:related}

As mentioned in the introduction, the vast majority of Internet of Things research to date has been focused on smart city and home environments \cite{atzori2010survey}. Instrumentation of the natural environment using IoT technology has received less attention. The key practical differences here are the lack of wide-area connectivity, lack of mains power sources, physical locations that are hard to access, and often harsh environmental conditions. We survey the most closely related work in this area below. Our deployment is the first to deliver a holistic IoT solution for the natural environment, enabling real-time study of multiple dimensions of that environment simultaneously.

One of the first experimental testbeds of a multi-tiered architecture for habitat monitoring was proposed by Cerpa et al. \cite{cerpa2001HMA} in 2001. Using the mote technology developed by Kris Pister at UC Berkeley \cite{kahn1999NCC}, the authors introduced the principles of building long-lived wireless sensor networks, adaptive self-configuration and its application to habitat monitoring. Their goal was to integrate these building blocks into a deployable system.

Researchers at the University of California, Berkeley, in collaboration with biologists at the College of the Atlantic, deployed the first documented tiered wireless sensor network architecture for habitat monitoring, such as detecting the presence of birds, on Great Duck Island (GDI), Maine \cite{mainwaring2002WSN}. They deployed two networks of a total number of about 150 nodes in both single-hop and multi-hop configurations. The initial deployment comprised 32 nodes using UC Berkeley motes, including temperature, photoresistor, barometric pressure, humidity, and infrared sensors. Two years later, the researchers analysed a rich set of data captured from these sensors from a systems perspective regarding the lifetime of nodes and network performance, reliability, routing and yield \cite{szewczyk2004ALS}.

Researchers from Princeton University deployed a 30-node wireless sensor network with mobile base stations called ZebraNet at the Mpala Research Center in central Kenya which aimed at monitoring and tracking wildlife, particularly zebras, for biology researchers \cite{juang2002ECW}. The ZebraNet system included nodes (tracking collars carried by animals) operating as a peer-to-peer network for transferring data. The collar, working as a wireless computing device, comprised GPS, flash memory, wireless transceiver and a CPU. ZebraNet was the first project to design protocols for sensor networks having a mobile base station.

The UCLA's Center for Embedded Network Sensing (CENS) deployed the Extensible Sensing System (ESS) for habitat sensing at the James Reserve in California. CENS was a multidisciplinary collaborative project to develop and deploy wireless sensing systems to monitor the biological, ecological, seismological and different environmental phenomena. The ESS had several components such as Directed Diffusion based communication software called Tiny Diffusion \cite{heidemann2003MDD}, Mica2 motes running TinyOS, and Compaq iPAQs. The sensors captured data regarding temperature, humidity, photosynthetically active radiation and infrared thermopiles for detecting animal proximity. The ESS was one of the first deployed sensor networks consisting of hundreds of nodes.

Researchers at the University of Southampton deployed an environmental sensor network called GlacsWeb \cite{martinez2005glacier} to monitor and study the dynamics of glaciers in hostile environments. The system comprised 9 nodes (probes) inserted in the glacier, a Base Station on the surface of glacier having a differential GPS and a Reference Station working as a gateway for transferring data. Each node was equipped with pressure, temperature and orientation (tilt) sensors for monitoring the melting behaviour of glaciers. GlacsWeb developed a low power sensor network supporting the transfer of sensor readings to a web accessible database.

Cardell-Oliver et al. \cite{oliver2004field} designed and implemented a reactive, event driven wireless sensor network for environmental monitoring of soil moisture. A novelty of their implementation is how their network reacts to extreme events such as intensive rainfall to send frequent readings and collect the most interesting data, reverting to less frequent collection in periods of lower rainfall. They used Mica2 433 MHz motes with MDA300 sensor boards comprising soil moisture and rainfall sampling nodes, a base node connected to a Superlite GSM gateway and routing nodes. They evaluated their network taking into account three performance parameters i.e. responsiveness, robustness and longevity of the network.

Researchers at UC Berkeley deployed a wireless sensor network, termed `Macroscope in the Redwoods', to monitor the complex spatio-temporal variations of the microclimate of a redwood tree \cite{tolle2005MR}. The sensor node was a Mica2Dot, a repackaging of the Mica2 manufactured by Crossbow. The mote consisted of an Atmel ATmega 128 microcontroller operating at 4 MHz, a 433 MHz radio and 512KB of flash memory. The measured biological parameters were temperature, relative humidity and light levels. Although Macroscope was deployed at a small scale it had the potential to measure the microclimate phenomena of a redwood tree at a spatio-temporal large scale. Further, it provided useful knowledge regarding deployment methodology and data analysis.

LOFAR-agro (Low Frequency Array) is the first deployment of a large scale sensor network in precision agriculture to protect potato crops from Phytophthora, a fungal disease, with close monitoring of the microclimate \cite{langendoen2006MLP}. In this project, around 100 nodes were deployed in a potato field to measure the temperature and relative humidity of the environment, known to be the main factors in causing the disease. The researchers experienced a range of unanticipated technical challenges due to poor design, not applying proper software engineering principles, semantic mismatches, programming bugs, and insufficient time for testing. 

Researchers at Harvard University in collaboration with researchers at the University of North Carolina and the University of New Hampshire deployed a sensor network on an active volcano in Volcan Reventador, northern Ecuador, to measure seismic and infrasonic signals \cite{allen2006volcano}. The network comprised 16 stations with seismic and acoustic sensors. Each node consisted of a Moteiv TMote Sky, a descendant of the UC Berkeley’s Mica mote, an 8-dBi 2.4GHz external omnidirectional antenna, a seismometer, a microphone and a custom hardware interface board. The network was deployed for more than three weeks and collected data from several events. The model was based on event triggering which detected 230 eruptions and captured about 107 Mbytes of data. 

In a collaborative project called `Life Under Your Feet', researchers at John Hopkins University with Microsoft deployed an experimental wireless sensor network in a Baltimore urban forest to monitor soil ecology \cite{musaloui2006life}. They used MicaZ from Crossbow Inc. with MTS101 data acquisition board comprising ambient light and temperature sensors. They attached Watermark soil moisture sensors and a soil thermistor to the board. They reported a mixture of previously known and new observations from the operation of their testbed, such as reprogramming the motes, sensor nodes' and calibration cost, the importance and challenge of low-level programming, the need for fast and efficient routing infrastructure and network design, and reliable delivery and high quality of data.

Bishop-Hurley et al. \cite{bishop-hurley2007VFA} designed and tested a prototype of a neck-collar-based virtual fencing system to monitor and control livestock movement and behaviour. They conducted an experiment on 25 steers at the Belmont Research Station, Australia, over a month. The electronics used in the experiment consisted of the Fleck WSN device in combination with a control board which could produce an electrical pulse. Two 6 volt batteries, placed in the collar’s compartments, were used to charge and power the Fleck. For radio communication, they used a 433 MHz wave flex whip antenna. Statistical analysis was done on observational and videotaped data to see how animals reacted to the stimuli. The experiment showed that the sensory cue has the potential to be used for monitoring livestock in large grazing systems.

Sensor networks have also been used to detect and monitor river flooding \cite{grace2008middleware} using hybrid local and remote modelling of observed conditions; this deployment particularly demonstrates the value of using heterogeneous hardware and of performing system adaptation based on locally observed conditions. Also in the natural environment, the GreenOrbs \cite{mo2009CCE} project is an extremely large deployment of sensor nodes in a forest environment to help understand canopy coverage over time.

SensorScope was an outdoor wireless sensor network deployed on top of a rock glacier in Switzerland for environmental monitoring in harsh weather conditions \cite{ingelrest2010SAS}. The researchers deployed 16 sensing stations on a 500x500m area for a couple of months to measure air temperature and humidity, surface temperature, incoming solar radiation, wind speed and direction, and precipitation. They opted for a Shockfish TinyNode sensor mote platform composed of a Texas Instruments MSP430 microcontroller running at 8MHz and a Semtech XE1205 radio transceiver. They designed and implemented the communication stack from scratch for their stations with TinyOS. SensorScope mainly aimed at providing a low-cost, efficient and reliable WSN with an emphasis on improved data collection techniques and to a great extent deployed successfully in outdoor harsh environments.

Corke et al. \cite{corke2010environmental} reported experiences learned from their wireless sensor networks deployed over six years to understand the natural and agricultural environments and the major challenges they pose in Australia. They developed and deployed sensor networks for different applications such as cattle monitoring, ground water and lake water quality monitoring, virtual fencing, and rainforest monitoring, highlighting the technological challenges they faced, which technology they used for each application and what lessons they learned from each deployment.




All of the above work suggests a strong need for Environmental Internet of Things developments, with high interest from multiple different domains of science. However, work to date remains relatively tightly focused on single dimensions of the environment, lacking a broader view that can fuse data across multiple scientific domains to build a holistic environmental perspective. Further to this, we note that a large proportion of the above works are technology-oriented and often fail to deeply involve scientists in the domain of study throughout the design process. In contrast to these two points, our work takes a multi-dimensional perspective on an environmental IoT deployment, and has deeply involved soil scientists, hydrologists and animal behaviour scientists from the first stages of design through to actual deployment and maintenance.



\section{Design and Deployment}
\label{sec:approach}

\subsection{Architecture and deployment overview}



Our environmental Internet of Things architecture spans elements in the field and in the cloud, providing an end-to-end solution that is exposed as an Internet service. In designing the system we had three goals: (i) to use off-the-shelf hardware and software as much as possible; (ii) to choose solutions that minimised the necessary amount of software development and system maintenance; and (iii) to provide a system that could be re-used by the environmental science team without expert help.

\begin{figure}[!t]
\includegraphics[width=3.2in,height=1.8in]{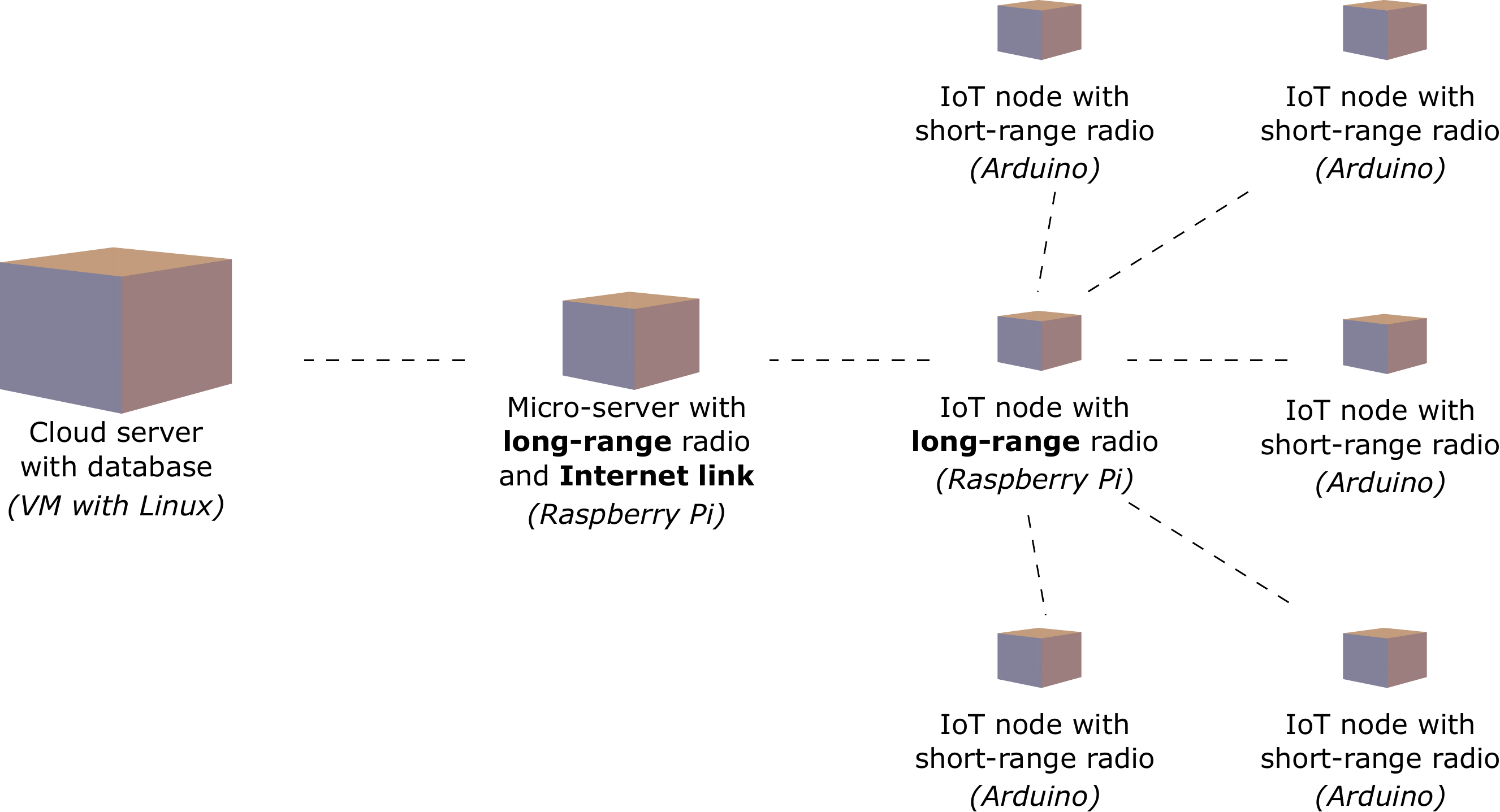}
\caption{Architecture of our IoT system. Further Internet services connect to the cloud server, providing scientific modelling or lightweight applications to visualise data.}
\label{fig:architecture}
\end{figure}

Our resulting architecture is shown in Fig.~\ref{fig:architecture}. In the field we use a micro-server as an Internet gateway (a Raspberry Pi) which is situated at a location with plentiful energy, either from a mains supply or from a large battery. This device is fitted with an Internet link technology, in our case 3G, as well as a long-range radio to communicate with the rest of the IoT deployment in the field. Our remaining IoT devices are then deployed into the actual field of study, which may be several kilometres away from the Raspberry Pi. One of these IoT devices ($I_A$) is itself equipped with a long-range radio and communicates with the Raspberry Pi; the remaining IoT devices have short-range radios and communicate only with $I_A$. This design has two major benefits: (i) the use of a separate Internet link node, which can be at a distant location, provides high flexibility because wide area coverage (i.e. 3G) may be sporadic or non-existent in the actual area of study; and (ii) the use of just one long-range radio `pair' reduces the cost of the overall deployment, as this higher class of radio is more expensive to buy and also draws more power, needing more expensive energy harvesting and storage technology at $I_A$. On the cloud side we then have one main server which receives data from the deployment, storing that data into a database. Other cloud servers (or end users) use web services to extract data from this machine and feed it into models or visualisation systems.

\begin{figure}[!t]
\includegraphics[width=3.6in]{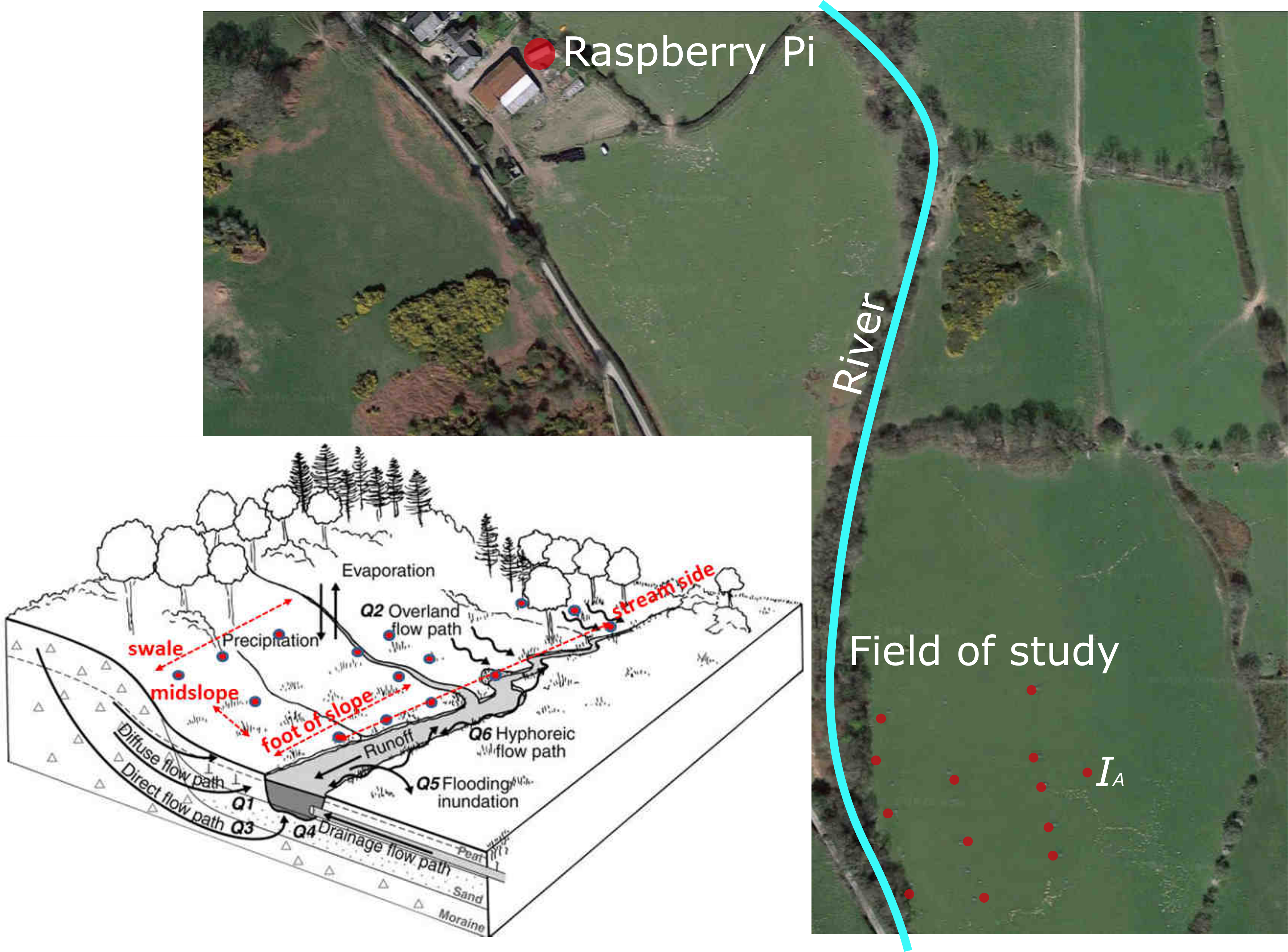}
\caption{Deployment location of our IoT system, showing a geographical view of the entire deployment as well as a detailed technical view of the field of study in hydrological terms. Mobile (livestock) nodes are not shown on the geographical view, but tracked animals are located within the treeline enclosing the area marked `field of study'. Satellite image \textcopyright ~Google Earth.}
\label{fig:field}
\end{figure}

In our particular case study, our system architecture was projected onto the deployment area shown in Fig.~\ref{fig:field}, providing both a geographical view and a detailed hydrology view. The Raspberry Pi is located in a small utility shed offered by a local farmer; our field of study is then $1.5km$ away from this location. The terrain between the Pi and the field of study also involves significant elevation changes and dense tree lines. Besides the long-range IoT node, which communicates with the Pi, we then have two types of node deployed in the field: fixed nodes measuring soil moisture levels, air temperature and humidity, and ground flow; and mobile nodes attached to livestock, measuring animal location and behaviour via GPS and accelerometers. Together these nodes provide an unprecedented real-time insight into multiple facets of the natural environment and their interrelationships. On the cloud side, data is fed into hydrological modeling software such as TopModel \cite{BEVEN1984119} and Jules; is fused with other available information including data from fixed weather stations and national meteorological services; and is also presented as a live visual feed for interested stakeholders.

In the remainder of this section we discuss the details of the nodes used in our deployment; the cloud infrastructure; and the user-facing elements of our architecture.



\subsection{Node technology}

As mentioned above, we use two types of sensing nodes –- soil sensing and livestock tracker nodes. The soil sensing nodes are used to take measurements related to the soil, whilst the livestock trackers are used to monitor the location of farm animals (specifically sheep during this project) in the area concerned. Both types of node have different physical requirements but need to be able to communicate with one another and so have a common link technology. In addition to this we use a long-range link node and a Raspberry Pi.

\begin{figure}[!t]
\includegraphics[width=3.5in,height=2.0in]{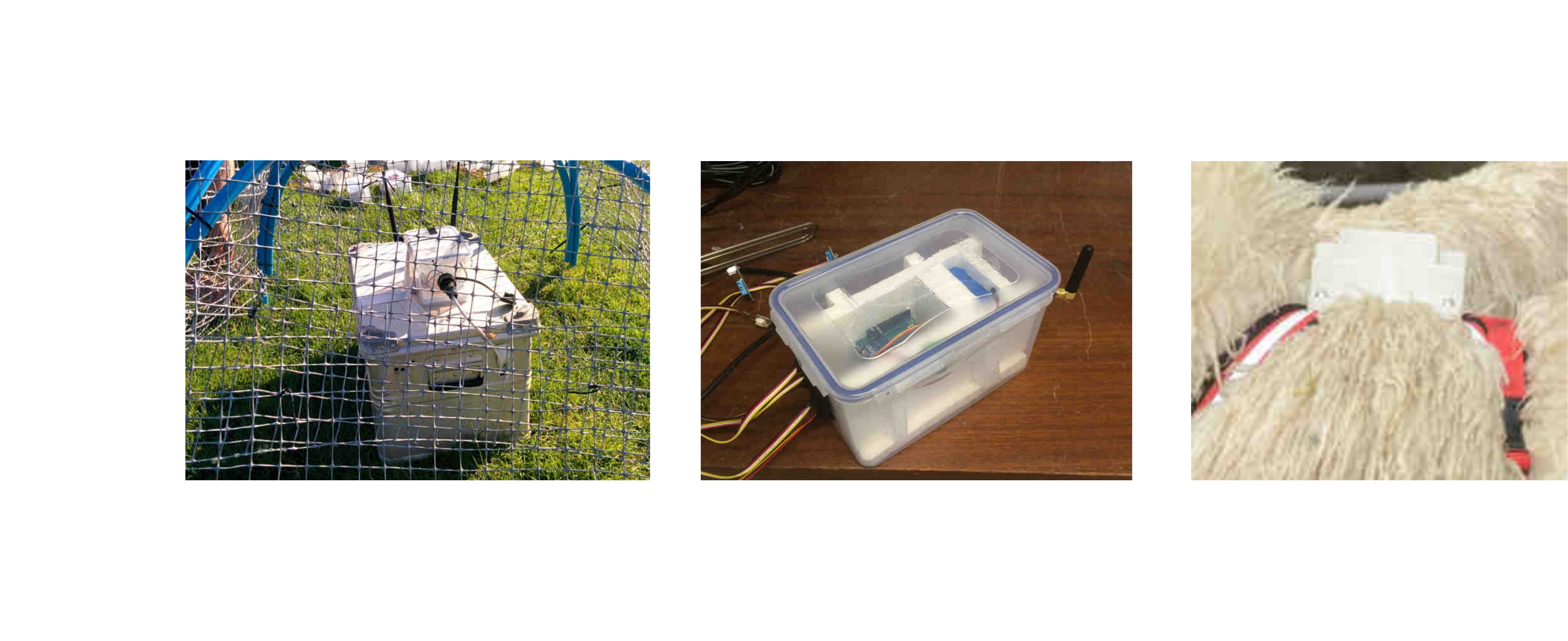}
\caption{IoT nodes (from left to right: relay, soil node, and livestock node on a sheep).}
\label{fig:node_imgs}
\end{figure}

In this section we present and justify our design choices in building these nodes, comparing with alternative existing technologies where available. Our overriding motivation was to develop a platform that was cheap and flexible. Images of each node type are shown in Fig.~\ref{fig:node_imgs}.

\subsubsection{Fixed nodes (soil)}

For both the fixed and mobile nodes we used Arduino technology as the basis of our solution. We made this choice after examining a wide range of IoT platforms including commercial off-the-shelf solutions. Our choice here was motivated by three major concerns: (i) the environmental scientists wanted a large range of sensing options, including custom-built sensory inputs; (ii) we needed a common technology that could support both fixed and mobile nodes working in the same network; and (iii) we were limited in our equipment budget. Looking at other technologies, IoT solutions tend either to be \textit{too specific} in their application domains, lacking the crucial configurability that we needed; lacked support for both fixed and mobile technology in a common network; or came in at a higher price point per node than we could afford. We therefore settled on Arduino-based nodes across our deployment. This was itself not without problems, however: we needed to develop more software ourselves and also had to deal with the relatively high power consumption of standard Arduino boards.

Our soil monitoring nodes cost \pounds150 each, including all sensing inputs, casings and power supplies. We chose to use the Arduino Mega2560, due to its relatively high processing power and larger collection of ready-built libraries. We used the Grove input extension board, allowing a high number of sensing inputs to be connected, with an Xbee Pro module for radio communication with other nodes. Each node also had two 7,800mAh batteries connected to it, providing sufficient power for almost a week. Each such battery had a built-in solar panel for trickle charging.

In the remainder of this section we discuss the issues of sensory inputs, power supplies and weather proofing for our soil nodes in particular.



\paragraph{Sensor inputs}

\begin{figure}[!t]
\includegraphics[width=3.0in]{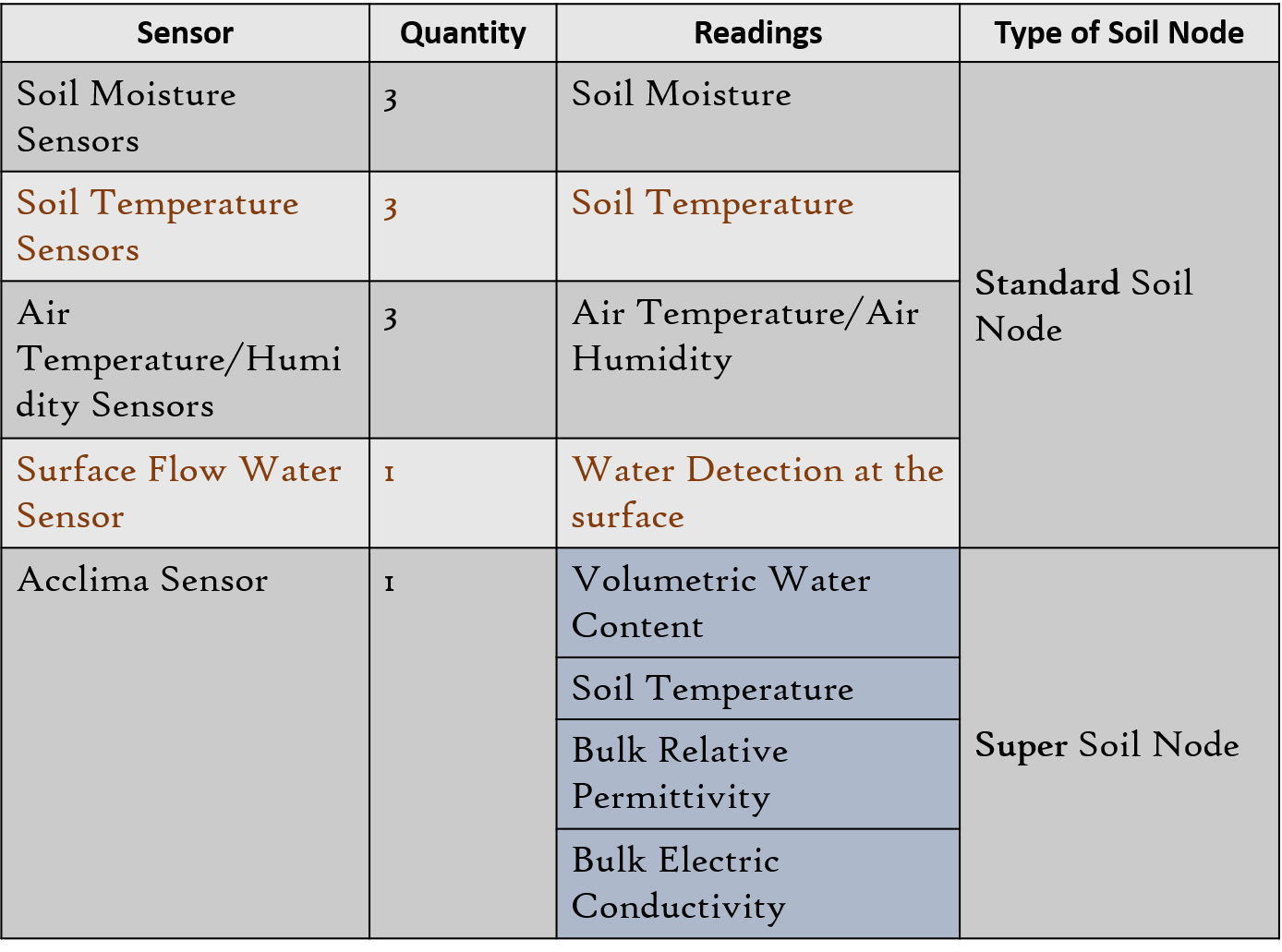}
\caption{Sensor inputs on our soil nodes, mixing a large quantity of cheap Grove sensors with a small number of high-quality Acclima sensors.}
\label{fig:soilsensors}
\end{figure}

Our sensor input strategy, guided by the environmental scientists on our team, was to deploy a large number of very cheap sensor inputs at most of our nodes, combined with a small number of high-resolution (but very expensive) sensors of known scientific quality. For our cheap sensors we used Grove hardware as listed in the upper part of Fig.~\ref{fig:soilsensors}. All of these are off-the-shelf units except for the physical design of the surface flow detector which was custom-built by the environmental science team.

As a high-quality counterpart to these sensors we used an Acclima soil moisture sensor (Digital TDT SDI-12). This sensor is considered to be one of the most accurate and stable soil moisture sensors available today. The Acclima sensor is more expensive than the Grove sensors, however, it is a preferred tool used by environmental scientists in order to measure different parameters in the soil. In typical (non-IoT) deployments, an Acclima is attached to a data logger unit and scientists periodically visit the site to download the data from this logger by physically attaching a USB drive. Instead we connected the Acclima directly to our Arduino board and wrote a simple driver to periodically take readings from it and send them through our IoT to the cloud.

In total, each soil node has 3 air temperature/air humidity sensors, 3 Grove soil moisture sensors, 3 soil temperature sensors, 1 surface flow sensor and (optionally) 1 Acclima soil moisture sensor. The reason behind replicating most of the Grove sensors was to get a more accurate and reliable value of the sensor readings in case one of them fails. Another reason was because the environmental scientists wanted to obtain a range of values at different spatial points for comparison (for example the air temperature sensors were mounted at three different heights). Moreover, although Acclima is a preferred sensor to Grove by the environmental scientists, the scientists wanted to experiment with the latter as a cheap alternative -- as the Acclima sensor is comparatively expensive. Hence, the experiment was carried out using the configuration of Grove sensors in order to see if such sensors do provide equally reliable readings to the Acclima sensor. The readings that have been compared between Grove and Acclima are mainly the soil temperature and soil moisture readings. As for the remaining Grove readings such as surface flow, air temperature and air humidity, these have been included in the experiment to get additional readings related to the soil and air. This has also been possible due to the fact that the Grove manufacturers provide a vast array of sensors which can easily be used and interfaced with the Arduino board.

A major question of the environmental scientists was whether the cost/accuracy tradeoff was a good one -- i.e., despite usually wanting to have the highest quality measurement equipment available, in this case they were willing to compromise on quality for quantity of real-time data points, as long as we still had some high-quality inputs in the field as a known reference point.

\paragraph{Power supply}

Powering the Arduino Mega2560 board, with all of its attached sensors, proved to be a major challenge. We found that the most reliable way to power the system was via its barrel connector, rather than the USB line. However, with a recommended voltage range of 7-12V via this connector, we required very high power batteries to achieve a sensible lifetime.

To help understand this lifetime issue, we measured the current draw of a complete sensor node, which was found to be 130mA when active and 45mA when in low-power sleep mode. Both values are extremely high compared to more bespoke sensor node solutions that use chips such as the TI MSP430. Additionally, the power draw is actually dominated by the Arduino board itself, which takes 90mA in active mode and 30mA in sleep mode. Using aggressive power cycling, if we assume that a soil node sleeps for 5 minutes and wakes for 5 seconds to send data then we can say that the node is awake for 1\% of the time and asleep for the rest. With this in mind we can calculate an average current draw of 46mA with 5 minutes of sleep, and with a 7,800mAh battery we achieve 7800/46 x 0.7 = 118 hours, or just under 5 days.

To meet the voltage input requirements and extend lifetime, each node was equipped with three 7,800mAh lithium ion batteries, offering 10 days of operations. For the environmental science team, this operating lifetime was around the same amount of time for which they would have regularly visited the field of study to take data from data loggers, and so having this as a worst-case battery replacement interval was acceptable. Operationally, the environmental science team kept a second set of batteries charged at their lab so that they could quickly replace any drained batteries in the field.

\paragraph{Weather proofing and protection}

Each sensor node was housed in a 1.9L plastic box enclosure, with space for batteries inside. The boxes were internally packed out with polystyrene to prevent movement of the components inside, and dessicant sachets were included in each node to counter build-up of condensation that could damage the electronics. For each external sensor, holes were drilled in the enclosure through which to pass the cables; these holes were then sealed with adhesive. Each node included a 4'' antenna (RN-SMA4-RP, Microchip) mounted externally using the RPSMA connector’s (MC000800, Multicomp) nut and a sealing ring. 

The individual Grove sensors, being generally designed for use in protected environments, were modified for field deployment. All were fitted with 50cm cables to the Arduino. Grove temperature sensors (Seeed Studio, 101020015) were fully enclosed in adhesive-lined heat-shrinkable tubing to include the cable termination. Hot melt glue was injected into the top of the tube prior to shrinking around the cable, creating a rigid section and protecting the connection. Three sensors per node were installed as soil-temperature replicates at 5cm depth. Grove moisture sensors (Seeed Studio, 101020008) and Grove water sensors (Seeed Studio 101020018) were similarly enclosed but leaving the probes bare. Three soil moisture sensors per node were installed in the soil surface. Each node also had one Grove water sensor which was used to prepare a prototype overland-flow detector. A plastic box was used to exclude rain from the sensor, with a wide cutout on the base of the upslope side and a drain on the other. Inside the box a V-shaped funnel was constructed in plastic to direct any overland flow over the sensor and out of the drain. Water flowing through the bottom of the V-shape would pass over the water sensor, completing the electrical connection to send a signal to the Arduino indicating over-land water flow. The Grove temperature and humidity sensors (Seeed Studio, 101020011) were shielded from rain under an upturned plastic cup. For each node, two sensors were installed at 35cm above the soil surface and one sensor was installed at 5cm above the soil surface.

The entire soil node assembly was placed inside a steel mesh dome to protect it from curious animals -- these domes were already at the field deployment site and had been well-tested by the environmental science teams in their pre-IoT data logger deployments.

\subsubsection{Mobile nodes (livestock)}

Our livestock nodes were designed to inform the environmental scientists of the location of animals and to correlate this with rainfall events that could lead to over-land flow causing contamination of the nearby river. We again based the design of this node on Arduino technology, this time using an Arduino Pro Mini as the main board, connected to a GPS module, a 3-axis accelerometer and an Xbee radio module. Our choice of Arduino in this case was entirely down to cost: commercial off-the-shelf solutions for animal tracking cost over \pounds1,000 per unit, whereas our solution cost just \pounds100. However, due to the need to minimise the scale of our unit to make it sufficiently lightweight to attach to an animal, we needed to hand-craft the design of this node much more closely, involving detailed electronic design of the board connecting the individual components together and also 3D-printing of the enclosure.

\paragraph{Sensor inputs}

As animal tracking is, in general, an area of scientific study that requires less fine detail than soil science, we had fewer technological constraints here coming from the requirements of the animal behaviour scientists. However, this was still a highly challenging design due to the interaction of technology with animals. We considered a large range of options for sensing animal location, from RFID tags on the animals with readers positioned around the field of study, to GPS units that would require more power and so a heavier overall node for animals to carry, and the use of web cameras sited at high points around the field of study (e.g. in trees) combined with image processing in the cloud as an approach that would completely avoid attaching equipment to animals. We settled on the GPS option as it would provide the most flexibility in what we could potentially do with our data streams, including the easy identification of individual animal behaviour. We chose to augment this with accelerometers as a way to potentially understand behaviour in a more detailed way.


\paragraph{Power supply}

Our animal tracking nodes were powered with a 10,000mAh battery, designed to be replaced once per week. Similarly to the soil nodes, we used aggressive power cycling with nodes in sleep mode most of the time. However, the amount of time that the animal tracker nodes needed to be awake was significantly extended by the duration of time that the GPS unit took to start up and then acquire a high-quality GPS fix.

\paragraph{Weather proofing}

The casing design for the animal tracking nodes was complicated by the fact that they needed to be both weather-proof and physically suitable to the animals without causing irritation. To provide a very close fitting enclosure for our electronics we 3D-printed the casing and then designed a harness to secure the device to livestock (in this case sheep). After several trials we settled on a harness made of soft but durable material which secured the sensor node on the top of the animal's back.

\subsubsection{Link nodes}

Link nodes are used to collect data received from our sensor nodes and deliver them to the cloud. In our IoT system, we have two kinds of link node: a long-range relay and a gateway node.

\paragraph{Relay Node}

The relay node is a Raspberry Pi model A+, sited in the field of study. We note that our initial network design used an Arduino as the relay; however it quickly became clear that the Arduino's relatively low-power CPU was unable to cope with forwarding the volume of data from the rest of the IoT nodes in the field. The relay node serves to receive data from the sensor nodes in the field and send them to the gateway node via a long-range local area radio transmitter. This design gives us much higher flexibility in the physical locations of the field of study and the (Internet link) gateway node. The relay node receives data from both the soil nodes and the livestock tracking nodes, being in local radio range of both kinds of node. The location of the relay was carefully chosen to maximise its radio signal propagation profile both in terms of the other IoT nodes in the field of study and the gateway node half a kilometre away.

The relay node is equipped with two different radios: a short-range Xbee and a long-range Xbee, each with their own antenna. The first Xbee (Xbee1) is used to receive data from soil/livestock nodes. The second Xbee (Xbee2) is then used to send data from the relay to the gateway. As we were aware that various weather and other local interference effects could disrupt the communication between the relay and the gateway, we implemented a simple store-and-forward system at the relay. Each data packet received by the gateway node from the relay is acknowledged, and any unacknowledged packets are stored locally at the relay node (using a TinyDB database) until their reception is confirmed by the gateway.

The relay node (Raspberry Pi A+) was powered via its USB socket. To facilitate this a DC/DC regulator 12V/5V (CPT, CPT-UL-1) was installed inside the enclosure. The output was attached to a USB charging cable, and the input to the bucanneer connector. This enabled a 12V 200 Ah deep cycle battery to be connected externally. This could be maintained with a solar panel but this was unnecessary due to high capacity and the availability of these batteries in our stores. 


\paragraph{Gateway Node}

The gateway node is a Raspberry Pi model B, and works in a similar way to the relay node, using a store-and-forward approach to push data into the cloud. Specifically, whenever a data packet is received from the relay node, the gateway node stores that data in a local TinyDB database. The gateway then periodically opens a connection to our cloud server (via a 3G cellular connection) and sends all data from its TinyDB database that has not yet been acknowledged as received by the cloud server.

We use TinyDB databases on the relay and gateway nodes as they are extremely lightweight, occupying minimal storage space on the relatively limited Raspberry Pi secondary storage. The location for the gateway provided 3G signal coverage and access to mains power. Therefore the gateway was powered via USB using a mains adapter plug for continuous power.


\subsection{Cloud Infrastructure}

The cloud side of our IoT architecture is relatively simple, using just one server as a database in which to store all data received from the deployment -- data which arrives as periodic packages from the gateway node over its 3G link. Other cloud resources can then access this server to use the data for specific purposes, such as scientific modelling or visualisations. We chose to use MongoDB as our data storage technology due to its schema-free design that is highly flexible to different kinds of data -- as is highly likely to be the case when using many different sensor modalities in the way that we have. Modelling and visualisation services are run on elastic cloud computing resources, scaling up the amount of processing power needed as their complexity increases.

Besides receiving and storing data, the main purpose of our cloud infrastructure is to act as a management service for the deployment. This allows nodes to be registered as they get deployed (including their position, sensing attributes and a logical group membership for each node), or for their details to be edited if they are later moved or changed, and also for the sampling rate of deployed nodes to be changed if desired. The latter is done using logical node groups, such that the sampling rate affects a selected group of nodes (e.g. `soil nodes' or `riverside soil nodes'). The management service also records the last time it heard any data from each node, and the last-reported battery level of each node, to aid in the day-to-day running of the deployment. As part of this service we have developed a mobile app through which the environmental scientists can add and update details about deployed nodes as they make changes in the field of study.

While this architecture is sufficient for our current deployment, in future we would explore more efficient solutions for processing and data distribution, particularly for modelling and visualisation services -- such as allowing some data processing to occur on the database server itself to avoid shipping large quantities of data to other locations.

\subsection{User interface and data presentation}

An important part of providing a truly end-to-end IoT solution is the user interface to the system. In this respect we have developed a significant `dashboard' for users to explore. We have also begun a semantic annotation initiative to connect our data to the semantic web for wider usability. We describe both of these elements below.

\paragraph{Dashboard Development}

Our dashboard application is an important part of `in the wild' research, gaining insights into what it means to deploy in the natural environment with all the complexities that this brings (different geographies, exposure to the elements, extreme events). The dashboard application makes use of the data streams collected through the IoT infrastructure that have been stored on our MongoDB database. The purpose of this dashboard is to enable the environmental scientists to quickly make sense of the data collected, informing them of the status of the sensed environment, of potential related hazards detected through the system, and providing room for decision making.

\begin{figure}[!t]
\includegraphics[width=3.2in, height=2.2in]{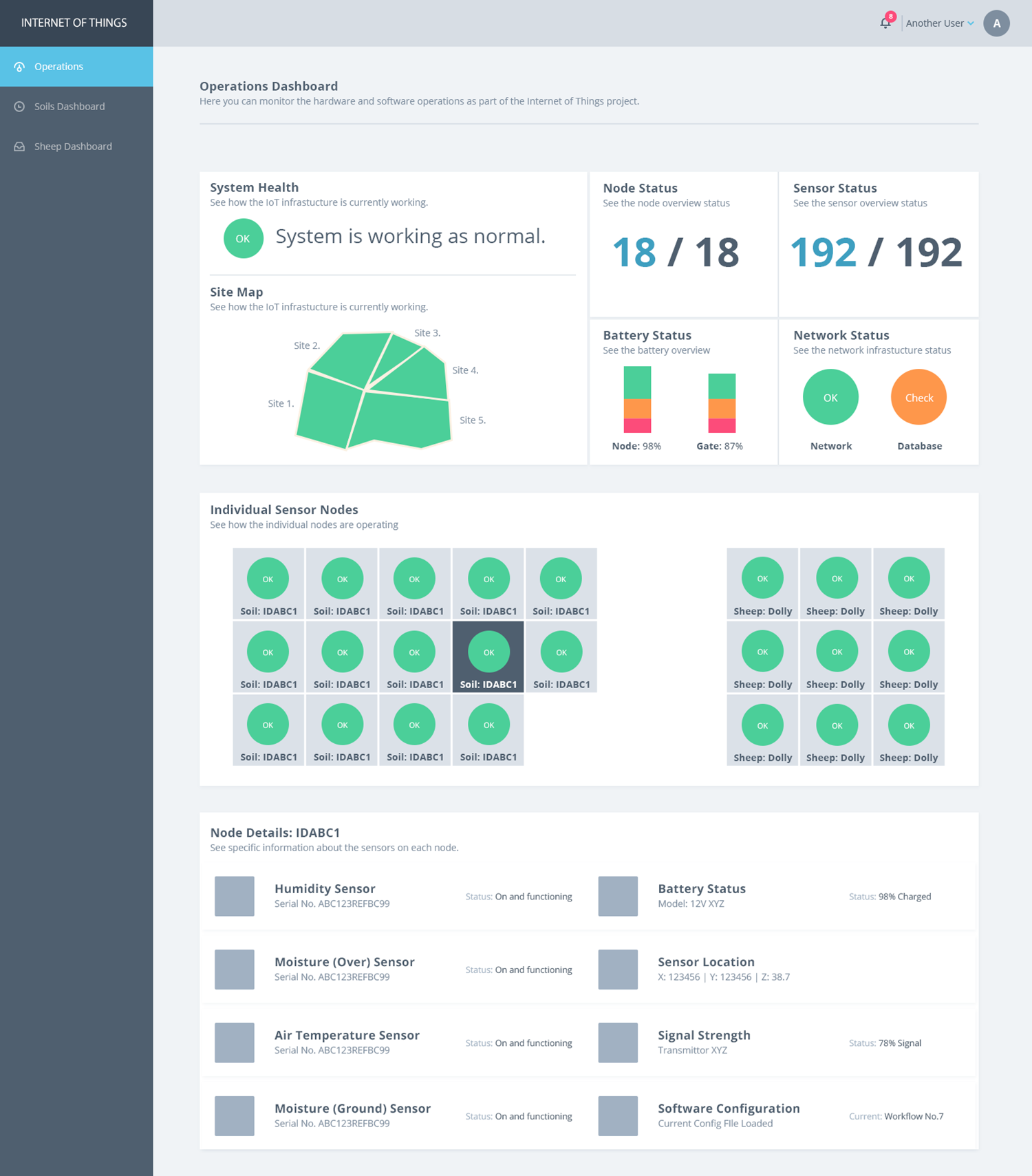}
\caption{Env. IoT Dashboard Application}
\label{fig:dashboard1}
\end{figure}

The data in the MongoDB comprises readings from the soil sensor nodes and the sheep trackers, all deployed in the Hiraethlyn field, as mentioned earlier. Whilst the soil sensor readings contribute towards environmental parameters such as soil temperature, soil moisture, air humidity/air temperature, and surface flow, on the other hand, the sheep trackers provide the GPS locations of sheep in the said area. A user-driven application has been developed in order to present the readings in a dashboard-style, as shown in Fig.~\ref{fig:dashboard1}.

\begin{figure}[!t]
\includegraphics[width=3.2in, height=2.2in]{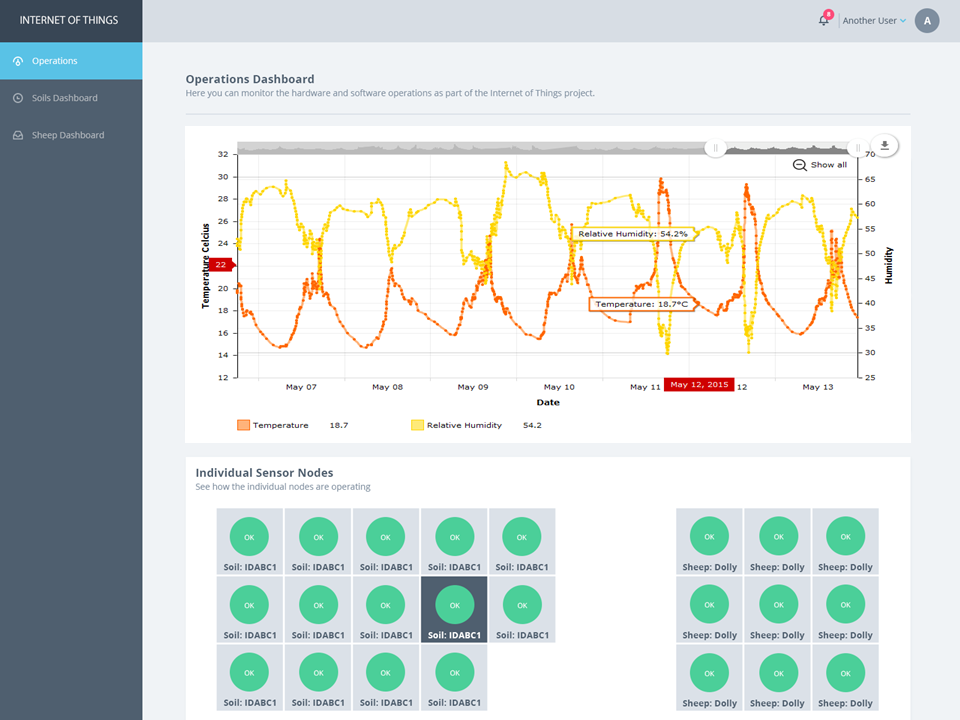}
\caption{Readings for one soil node}
\label{fig:dashboard2}
\end{figure}

The dashboard consists of different windows, namely an operations dashboard, a soil dashboard and a sheep dashboard, as indicated in the left pane of Fig.~\ref{fig:dashboard1}. The operations dashboard is the main window and shows the number of sites where the IoT infrastructure has been deployed, the number of soil nodes and sheep trackers used in each site, the status of the nodes (whether they are working), and the kinds of readings being received from each type of node. The application also enables the user to select a node and to analyse the readings obtained from this particular node, as shown in Fig.~\ref{fig:dashboard2}. This figure shows the readings from the air temperature/humidity sensors from one soil node over a one-week period. Each of these windows is populated using pluggable `widgets' which can be individually developed to add to the dashboard. This is another key element of configurability to support a highly flexible set of current and future sensor modalities.


\paragraph{Semantic Data Enrichment}

Whilst the dashboard application development focused on presenting the `raw data' collected through our IoT network in a significant way, we are also venturing further with this development by exploring how to semantically enhance our system. The MongoDB data can be used to plot the sensor readings obtained over a time period on the dashboard, thus providing a medium to the environmental scientists to carry out in-depth analyses over these sets of data. However, the data stored on MongoDB does not provide any other additional information to the scientists, other than the sensor readings, that may improve the analysis of the sensor readings. This entails information such as the types of sensors used to collect the readings and the measurement capabilities of such sensors, the related conditions under which the readings have been taken etc.

The enrichment of the sensing data collected with such metadata can be made possible through the use of semantic technologies, namely ontologies. In general terms, an ontology is defined as ``a formal, explicit specification of a shared conceptualisation'', as proposed by Tom Gruber\footnote{http://semanticweb.org/wiki/Ontology.html}. It is a data model that enables us to define a context or domain through a set of formal, interrelated vocabulary terms.

Therefore, we have created an ontology to capture the features of the IoT network, partly based on extending several existing vocabularies such as SSN/DUL\footnote{https://www.w3.org/2005/Incubator/ssn/wiki/DUL\_ssn} to describe the kinds of sensors used and their measurement capabilities, the Time ontology\footnote{https://www.w3.org/TR/owl-time/} to describe the timestamp of the values read, MUO\footnote{http://www.semantic-web-journal.net/sites/default/files/swj177\_7.pdf} to describe the units of the values read by the sensors, and Geo/Geosparql\footnote{http://www.opengeospatial.org/standards/geosparql} to describe the location of the deployment site.

The ontology also enables us to reason more deeply over the data model and to infer additional knowledge that pertains to the domain of interest. At we high level we therefore hope to be able to answer complex and multi-dimensional queries of the form: can we correlate the soil moisture readings to the type of weather/or rainfall that occurred during that time period?; can we infer that the nearby river was polluted due to sheep wandering in the Hiraethlyn region when there was heavy rainfall?; did the rainfall trigger a run-off of water (with pollutants/FIO) due to the soil having reached a high level of saturation?

The exact way in which these queries can be expressed, particularly in a form that suits domain scientists, remains an open research topic. The addition and use of semantic annotations to data is also an important part of provenance, particularly in the hard science use case that we have here; we return to the issue of provenance more broadly in Sec.~\ref{sec:discussion}.



sec\section{Deployment experiences of an environmental IoT}
\label{sec:evaluation}

\begin{figure*}[ht]
\centering
\includegraphics[width=6in]{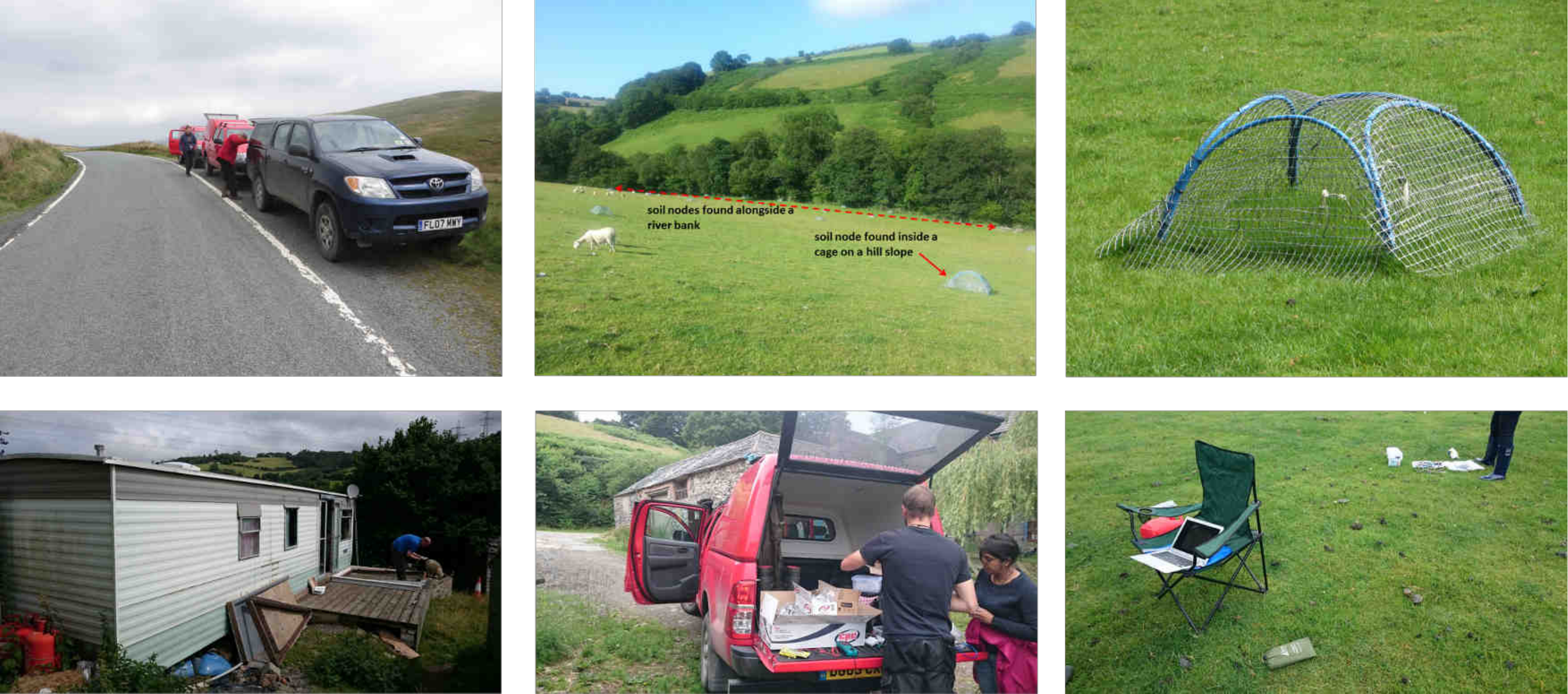}
\caption{Images from our deployment site. Clockwise from top left: initial site reconnaissance with environmental scientists; the site as we initially found it; a protective dome over a soil sensing node; configuring the relay node; constructing hardware at the site; and the location of the Internet gateway node.}
\label{fig:experience}
\end{figure*}

In this section we report on some of the main deployment and operations stage of the project; a collection of images from this experience is shown in Fig.~\ref{fig:experience}. We first note that our deployment timing coincided (by chance) with one of the most severe storm and flooding periods in UK history, during November-December 2015, creating extraordinarily challenging conditions for operating technology in the field. The most significant problems caused by this were lower than anticipated solar charging levels and terminal water damage to some nodes. Nevertheless, our deployment was broadly successful, resulting in large volumes of real-time data collection from both the soil nodes and the animal trackers.

\paragraph{Design of the soil node}
The basic enclosures were sufficient to exclude rain from the relay and soil nodes. There was some difficulty in excluding heavy rain from the hardware during soil node battery changes, which could be resolved with an external battery connected via waterproof connector, avoiding the need to open the housing of the IoT node itself. Additionally, soil nodes near to the watercourse in particular were damaged by water ingress when submerged by flooding.

\paragraph{Design of the animal tracking node}
This node performed well, and was unaffected by rain issues, though some build-up of condensation inside the node housing was observed, likely a particular problem as these nodes were in prolonged contact with warm animals and an unusually cold and wet environment. Reports from the local farmers indicated that the sheep to which the nodes were attached appeared behaviourally unaffected by the deployment.

\paragraph{Battery Usage (interval for changing batteries)}
The runtime of all of the nodes was shorter than anticipated, requiring attendance at the field site at 5 day intervals by the environmental science team to attend to nodes which had ceased reporting. We consider four reasons for this. 1) Additional consumption of power by the attached sensors may have been greater than estimated. 2) The 3.6V battery packs, connected in series, each include an internal circuit protecting against over-charge and discharge. Any discrepancy in charge between the packs comprising a unit would result in one component pack reaching upper / lower protection thresholds ahead of others, interrupting the circuit for all component packs. With this experience we consider that 12V, deep cycle, sealed lead acid batteries would offer a more cost-effective and suitable power source despite somewhat larger and heavier format. These could be connected externally and maintained by a solar panel depending on node situation. 3) Later in the project we detected a problem with the sleep / wake cycle of the XBee, which resulted in nodes ceasing to report and which may have been mistaken for a discharged battery. The frequency of this occurrence during field deployment is not clear but in hindsight the observation that nodes could often be restarted by cycling the power, rather than replacing the battery, and observation that a healthy voltage was often measured in batteries recovered for charging, suggests that this error may account for a significant number of early drop-outs. 4) Exceptionally poor weather limited solar recharge effectiveness across the deployment. In addition, we note that we were never able to operate the accelerometers on the animal tracker nodes as the volume of radio traffic generated was too high a drain on their batteries. Independent trials suggested that on-node classification algorithms could have been used to process data locally before sending a simple summary status of the animal such as `standing', `sleeping' etc. As a general comment on our platform, we found that the Arduino hardware that we chose had a surprisingly high power draw even in sleep mode; a modified solution that uses a low-power micro-controller to wake up the main Arduino chip when needed could have significantly helped with this but was beyond our capability to implement in the time constraints that we had.

\begin{figure}[!t]
\includegraphics[width=2.4in]{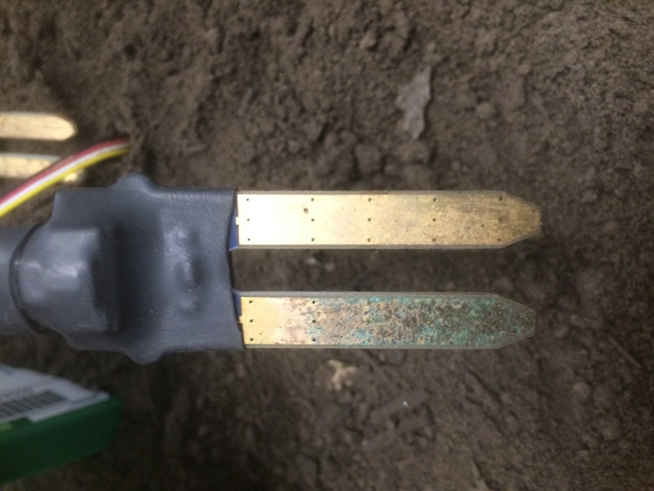}
\caption{Corroded Soil Sensor}
\label{fig:corrodedsoilsensor}
\end{figure}

An early guide from the environmental science team indicated a desire to try using a large number of cheap sensors rather than fewer more expensive sensors. The motivation behind this was to reduce deployment cost, sacrificing data quality for quantity in some cases, while still maintaining selected high-quality sensors in the field as calibration points. For our particular deployment conditions (particularly the extreme weather) this does not appear to have paid off, with a high frequency of anomalies in sensor data being apparent. For the prototype overland flow detector and the air temperature / humidity sensors, problems with the constructed housings are likely to have contributed to anomalous data. Overland flow detectors are not commercially available due to difficulties in designing a reliable instrument and our prototype suffered problems with false-positives due to internal condensation and blocked drainage. Subsequently the Grove water sensor, incorporated to detect the presence of water in the channel, became corroded due to the relatively cheap copper alloy used in its construction. Research-grade air temperature and humidity sensors are usually enclosed in radiation shields to minimise heating by solar radiation and direct exposure to precipitation. These may be passively or actively aspirated. These shields cost much more than our chosen sensors. We used white plastic cups as shields which excluded precipitation but were not vented and may have trapped heat and condensation. The Grove soil moisture sensors also generated anomalous data due to the same copper corrosion noted above; while we were able to protect the control circuitry we could not defend against electrolytic corrosion of the moisture probes. Fig.~\ref{fig:corrodedsoilsensor} shows a corroded Grove soil moisture sensor. We stress that the Grove soil moisture sensors are not intended for this kind of `heavy duty' use with long-term submersion in standing water -- had the weather been closer to general trends for this time of year, we would likely have had less problems in corrosion as most locations in the field of study are less moist. 

\paragraph{Calibration of the sensors (Grove vs. Acclima)}
As we had collected a significant volume of data from the deployment, at the end of the field deployment period we devised a lab-based experiment to seek a calibration of the Grove Soil Moisture sensor data against that of an Acclima TDT. We powered 4x nodes at a constant 12V using a laboratory power supply. We attached Acclima TDT sensors to the nodes and assigned each node to one of four 35 litre planters filled with topsoil. We installed the sensors as per the field deployment and placed the relay and gateway nearby. We then saturated the planters with water before drilling holes and leaving the planters to dry out to compare the drying curves revealed by the two soil moisture sensor types. This phase of our work (which is ongoing) was carried out without direct assistance from the computer science team and is a positive indication that we are able to reuse the field technology in different settings without expert knowledge. In future we hope to be able to redeploy the network in other fields of interest.

\paragraph{Simple network design}

Our final network design was effectively a simple star network with one long-range hop to the Internet gateway. This was the result of significant field testing of a range of radio technology at the deployment site. Our initial plan was to make use of a multi-hop data collection protocol, as is common in the literature, but field testing indicated that this would not be required. In fact we observed extremely good levels of point-to-point connectivity across long distances with treelines and other obstacles between nodes. As an anecdotal example of this, one of our nodes was transported in an off-road vehicle to a distant location for range testing; this node remained in contact with our base station for the entire journey despite being inside an enclosed steel vehicle. As a result we adopted our current simple network design with one node sitting at the top of the hill in the field of study, acting as a relay for all other nodes in the field. The relay then sends its data to a gateway node which sits in the barn at the site. If future deployments were to need a node dispersal over a larger physical radius (more than 2km apart) a more complex network design may then be required -- our most likely solution would be to adopt a design using further relay nodes as cluster-heads in more distant parts of the network, such that only the relays communicated over longer distances.
\section{Discussion}
\label{sec:discussion}

This section provides a more high level discussion of the contributions of the project, revisiting in particular the three key aspects of the project: the technological, scientific and methodological aspects. We look at each area in turn, reflecting on the key contributions and insights in each area.

\paragraph{Technological contributions} This project was by definition an `in the wild' deployment in a complex and remote area of North Wales and this dictated a lot of the design choices. In addition, the work was in support of real scientific enquiry and hence had to provide a complete and comprehensive solution from sensor to data portal. This resulted in a pragmatic and end-to-end architecture that achieves seamless integration across both Internet of Things and cloud domains. The pragmatism is most evident in the network architecture (as presented in Section \ref{sec:approach}). For example, we initially planned to deploy a mesh network architecture for added coverage and resilience. With the range and effectiveness of the XBee Pro radios, this was however deemed unnecessary and instead we opted for the simpler point to point architecture, supplemented by the additional bridging nodes with long range radios (effectively better antennae). With careful placement of the bridging node(s) this solution proved successful. We also initially planned to achieve end-to-end integration through IP, as advocated by many researchers in the IoT community \cite{colitti2011rest,ko11beyond}. This however was not implemented and instead we opted for a more `systems-of-systems' architecture, reasoning about individual systems and the boundaries between them. This again proved to be a successful strategy and allowed us to optimise the protocols in the IoT deployment site for their particular purpose. The micro-server was then particularly important in acting as an Internet gateway in our systems-of-systems approach. Further investigation of systems-of-systems architectures is being carried out in the Dionasys project, including further investigation of this application domain \cite{blair2015HTS}.

The work highlighted the importance of flexible and adaptive placement of computation in the resultant systems of systems approach. Our experiences indicate that it is far too simplistic to bring the data back to the cloud for processing, and this is particularly important in this domain. For example, in many circumstances, the data is rather static and predictable but then at times of extreme events, such as storms/ heavy rainfall, then the data becomes critical to understand flooding behaviours of potential water quality issues. It is also too draining in resources (bandwidth, battery) to communicate all data and also sensors such as our accelerometers (on the sheep) generated vast volumes of data. These insights have led to future work in: i) the dynamic enhancement of the micro-server to be an edge device looking at constructing a micro-cloud out of a number of Raspberry Pis, ii) strategies looking at the dynamic placement of computation on the resultant complex distributed environment (incl. the pushing of behaviour to end or edge devices, iii) self-organisation strategies for the above based on machine learning.

The combination of IoT technology and cloud computing is powerful -- with IoT providing a source of rich, ubiquitous data coupled with an ability to intervene in the real world through means of actuation, and cloud computing offering massive and elastic capacity in terms of storage and processing, coupled with an associated service architecture to interpret, manipulate and visualise this data. There is however one missing piece -- data science -- and it is clear from our work that more effort is required to make sense of the rich volumes of data generated by an Environmental IoT. It is also clear that this domain represents different challenges for data science than many areas of big data; the data in this area tends to be a larger collection of heterogeneous data elements requiring complex analysis around, for example, inter-dependency. Researchers refer to this as the `long tail of science' \cite{heidorn2008shedding}. As a further follow up, we are investigating this eternal braid of the three synergistic technologies in the understanding, mitigating and adapting to climate change.

\begin{figure}[!t]
\includegraphics[width=2.7in]{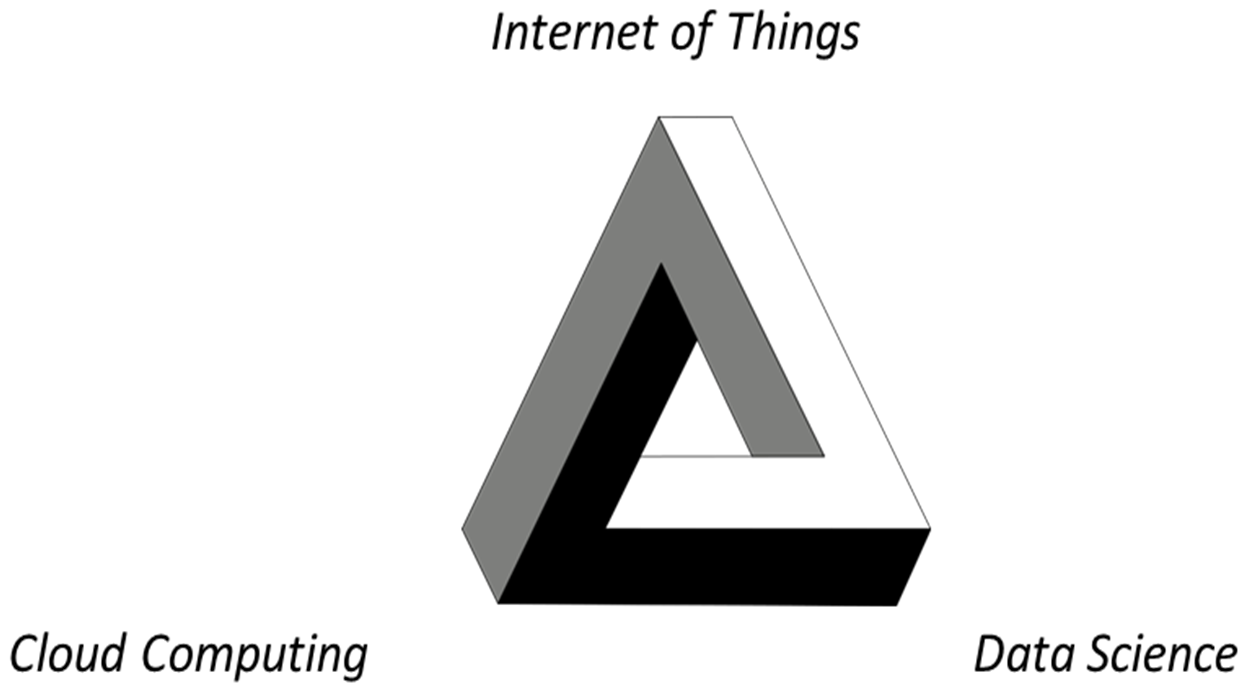}
\caption{The complementarity of IoT, cloud and data science is a key field of future work for integrative research efforts -- each needing to understand the role and capabilities of the others.}
\label{fig:datascience}
\end{figure}

\paragraph{Scientific contributions} Arguably the biggest success of the project was that we provided a working and fully functional Environmental IoT that enabled new modes of scientific discovery. Key to this was the deployment of a wide range of sensors covering hydrology, soil science and animal behaviour. This enabled a new kind of integrative scientific enquiry and also, by bringing the data together, this also brought the associated scientific communities together to collaboratively understand the inter-dependencies in the complex environmental catchment under consideration. This acts as a beacon for future integrative and collaborative science in this area. The examination of this multi-faceted data across different geographic areas and (importantly) at different scales also enables this broader and deeper understanding of a catchment in all its complexities. The more obvious but equally profound contribution is providing real-time data to scientists. As mentioned in related work, environmental and earth sciences are overly reliant on data logging which require trips out to (often remote and inaccessible) locations to download data. The accessibility of rich, real-time data therefore represents a real paradigm shift for this area and a major opportunity to support improved response to issues such as, for example, extreme events and natural disasters.

One key issue that emerged was the need to understand the semantics of the data, most notably related to the provenance of data (what type of sensors, their perceived accuracy, and so on). For example, we deployed a range of soil sensors covering a range of prices and qualities. This is important as scientists are interested in what can be achieved from dense deployment of very cheap sensors but equally need more accurate technologies for calibration. Hybrid technologies are therefore important. We have thus devoted significant effort to the resultant (semantic) data model, covering provenance but also other meta-data descriptions and supporting more conceptual reasoning about the data. This data model is briefly mentioned in Section \ref{sec:approach}, and is also the subject of a sister paper currently under development.

A final important design decision was the provision of data in the cloud as a key means of enabling integration and collaboration, but also crucially in supporting open access to data (available through MongoDB). The importance of open data has already been recognised by many observers, most notably by the Royal Society \cite{royalsoc}. This is particularly important in the field of environmental and earth sciences as the communities respond to the requirements of a new kind of `big' science as required by the challenges of environmental change. Open data also encourages innovation around both uses of the data and (importantly) associated services. Access to the Environmental IoT data is available through the associated data portal, but the open API also encouraged consideration of a range of apps including a `MySheep' app, and also the potential for integration into the existing MySoils app \cite{Shelley2013}.

\paragraph{Methodological contributions} It is often the case that IoT developments are carried out by the computer science community on behalf of end users and this can result in technology for its own sake. In this case, it was crucial that the project was a truly equal collaboration between technologists and environmental scientists. This was necessary to allow the scientists to appreciate what is possible from the underlying technology, often to enable a broader vision to develop, and also for the technologists to fully understand what is needed to achieve a true paradigm shift in scientific understanding. This was achieved, representing arguably the biggest contribution to the project dwarfing, for example, any technological insights. The key to this was an agile development method, and the associated use of flexible technologies such as Arduinos and Raspberry Pis. While these technologies are not necessarily optimal in terms of final development, they are ideal for rapid prototyping and exploring a space -- and indeed as a basis for having a conversation between the different constituent disciplines involved. It was notable that within a short time the scientists in the project were adapting the initial designs to better match their needs and also to extend the functionality. In effect, the technological and methodological approaches enabled a conversation on the role of Internet of Things technology in the environment and there is also evidence emerging through follow-on projects that this has resulted in a deeper and more sustainable level of understanding.

\section{Conclusion}
\label{sec:conclusion}

This paper has examined the potential role of Internet of Things technology in supporting a deeper understanding of the natural environment, reporting in particular in the design and `in the wild' deployment of an exemplar Environmental IoT examining a range of environmental factors in a catchment in North Wales. In concluding the paper, we return to our central hypothesis:

\textit{``A combination of IoT technology coupled with Cloud Computing enables a paradigm shift in our understanding and management of the natural environment, especially related to understanding ecosystem inter-dependencies, in times of unprecedented environmental change.''}

This is clearly a rather bold and open ended claim but nevertheless our experiences indicate that this is an area of huge potential and that our experiments have shown that new styles of scientific enquiry can be supported by such emerging technology. We also flag this as an important area of future development of IoT technology. While it is often the case that IoT deployments feel like a technology in search of a problem, it is absolutely the case that in the environmental and earth sciences it is the other way round -- a problem in search of a solution where at least part of the solution lies in the combination of cloud and IoT technologies, coupled with emerging techniques from data science as discussed above.

The one barrier is the lack of maturity of Internet of Things technology based partly on the lack of standardization in this domain and also the many open research questions to be resolved, for example around minimization of energy and self-management. We have outlined some key areas of future research following on from this work most notably on areas such as systems of systems architectures, the role of edge devices and dynamic placement for generally, and self-organisation in complex distributed systems, alongside the aforementioned work on data science methods for real-time environmental data.




\bibliography{references.bib}{}
\bibliographystyle{IEEEtran}
\end{document}